# Surface-plasmon opto-magnetic field enhancement for all-optical magnetization switching

A. Dutta[a,b*], A. V. Kildishev[a,b], V. M. Shalaev[a,b], A. Boltasseva[a,b], E. E. Marinero[a,b,c]
[a]School of Electrical and Computer Engineering, [b] Birck Nanotechnology Center, [c]School of Materials Engineering, Purdue University, West Lafayette IN, USA 47907

*Corresponding authors: dutta6@purdue.edu, eemarinero@purdue.edu

## ABSTRACT

The demand for faster magnetization switching speeds and lower energy consumption has driven the field of spintronics in recent years. The magnetic tunnel junction is the most developed spintronic memory device in which the magnetization of the storage layer is switched by spin-transfer-torque or spin-orbit torque interactions. Whereas these novel spin-torque interactions exemplify the potential of electron-spin-based devices and memory, the switching speed is limited to the ns regime by the precessional motion of the magnetization. All-optical magnetization switching, based on the inverse Faraday effect, has been shown to be an attractive method for achieving magnetization switching at ps speeds. Successful magnetization reversal in thin films has been demonstrated by using circularly polarized light. However, a method for all-optical switching of on-chip nanomagnets in high density memory modules has not been described. In this work we propose to use plasmonics, with CMOS compatible plasmonic materials, to achieve on-chip magnetization reversal in nanomagnets. Plasmonics allows light to be confined in dimensions much smaller than the diffraction limit of light. This in turn, yields higher localized electromagnetic field intensities. In this work, through simulations, we show that using localized surface plasmon resonances, it is possible to couple light to nanomagnets and achieve significantly higher opto-magnetic field values in comparison to free space light excitation.

## 1. INTRODUCTION

Following the experimental demonstration of ultrafast (few hundreds of fs) demagnetization in 1996 utilizing optical pulses [1], the magnetization switching dynamics in these unprecedented time scales has been extensively investigated both theoretically and experimentally [2]. We note that magnetization reversal speed induced by magnetic fields or spin polarized currents is limited by the precessional motion of the magnetization. This limitation hinders the implementation of current spintronic devices in beyond-CMOS electronics and in particular, for on-chip integration of memory and logic spintronic-based devices. In contrast to CMOS devices that switch in ps time scales, state of the art spin-transfer-torque magnetic tunnel junctions (STT-MTJ), typically switch in the ns regime. Furthermore, high current densities of the order of $10^7 - 10^8$ A/cm$^2$ are required, making them unattractive from an energy consumption perspective. In addition such high current densities can result in electrical breakdown of the MgO tunnel barrier, negatively impacting their reliability.

All-optical magnetization switching has been shown in recent years to overcome precession dynamics limits and 40-fs circularly polarized laser pulses have been employed to reversibly switch the magnetization by changing the light helicity in a wide range of magnetic materials [2, 3-6]. Whereas the full physical understanding underlying the ultrafast magnetization reversal remains incomplete, one plausible mechanism invokes the fact that circularly polarized laser pulses provide an effective magnetic field via the inverse Faraday effect. This opto-magnetic field, $H_{OM}$, is proportional to [E x E*] (E is the light electric field). $H_{OM}$ is largely responsible for the ultrafast magnetization reversal [5]. The magnitude of this field depends on the intensity of the incident light as given by the following equation:

$$H_{OM} = \beta \varepsilon_0 (E \otimes E^*) \quad (1)$$

where $\beta$ is the magneto-optical susceptibility of the magnetic material, $\varepsilon_0$ is the permittivity of free space and $E$ is the magnitude of electric-field of the electromagnetic wave inside the magnetic material. The magneto-optical susceptibility $\beta$ is calculated using the following equation:

$$\beta = \frac{\theta_F \lambda n}{\pi d M_0} \quad (2)$$

where $\theta_F$ is the Faraday rotation at wavelength $\lambda$ for a magnetic material layer with magnetization $M_0$ and real part of refractive index $n$. It is estimated that a laser pulse fluence of 1 mJ/cm$^2$ generates an opto-magnetic field of 5.2 T!! In [5], the authors utilize 20nm thick films of GdFeCo and show magnetic domain reversal in sub-picosecond timescales for fluences less than 4.5mJ/cm$^2$. For higher fluences, their results show the formation of multi-domain states as well as helicity-independent switching which implies that there is an upper bound to the maximum light intensity that can be used for all-optical magnetization reversal. Their experimental setup consisted of a pump pulse of photon energy 1.54 eV with 70μm diameter and pulse width of 100 fs and a probe of photon energy 1.94 eV with 300μm diameter and 100 fs pulse width.

All experiments and theoretical discussions regarding all-optical magnetization switching published thus far, relate to the direct interaction of the electric field of the incident free-space propagating light with the magnetization of the material. The use of gold antennas placed on the surface of SiN protective layers of TbFeCo thin film stacks for magnetization switching with linearly polarized light has been reported by Liu et al [7]. In this work, we study the effect on the opto-magnetic field of surface plasmons generated with circularly polarized light at the interface between magnetic materials and plasmon resonators. We find that the magnitude of the plasmon-generated opto-magnetic field, $H_{OM}$, is enhanced in comparison to that resulting from the direct photon-magnetic material coupling. In particular, we find very large enhancements (~10x) when the magnetic material is either a dielectric or is integrated within a dielectric structure rather than a conventional metallic thin film. The utilization of surface plasmons for all-optical switching proposed in this work, paves the way for on-chip integration of nanoscale photonic and spintronic devices for beyond-CMOS circuitry.

## 2. STRUCTURE AND MATERIALS

Plasmonics deals with the study of the oscillations of free electrons in a metal coupled to an electromagnetic field [8]. These oscillations, known as surface plasmons [9], are associated with wave-vectors much larger than that of free-space electromagnetic waves. Therefore, plasmonics enables the localization of electromagnetic energy to dimensions much smaller than the diffraction limit. We study the role of plasmonic resonators coupled to nanomagnet structures in the generation of opto-magnetic fields under illumination with circularly polarized light. In particular, through numerical simulations, we demonstrate that under identical laser fluences, significantly larger opto-magnetic fields are generated in the coupled plasmonic/magnetic nanostructures when compared to direct light excitation of the nanomagnet. Whereas we focus in this work on the switching of nanomagnets with perpendicular anisotropy (PMA), as they are the magnetic materials of choice for ultra-high density memory and logic devices [10], and therefore, our discussion pertains mainly the z-component of the opto-magnetic field generated by surface plasmons. However, our findings are readily applicable to in-plane magnets. Moreover, we also indicate a possibility where our designs would be useful for switching magnets engineered to exhibit tilted magnetization orientation [11].

Figure 1a shows a schematic diagram of the coupled magneto-plasmonic structure. It consists of a nanodisk stack comprising a plasmonic resonator (yellow), a nanomagnet (purple) and a capping layer (green) fabricated on an optically transparent substrate (white). With respect to the choice of materials for plasmonics, the use of gold [12] and silver [13] has dominated the literature. While these materials show excellent plasmonic properties, they suffer from drawbacks such as poor mechanical and thermal stability as well as incompatibility with CMOS processing. Consequently, over the past few years significant research efforts have been dedicated in the search of low-loss CMOS compatible metals that can substitute gold and silver [14,15]. In this regard, refractory transition metal nitrides have emerged as suitable candidates for the visible and near-infrared (NIR) electromagnetic spectrum [16]. In particular, titanium nitride exhibits optical properties comparable to those of gold, it is compatible with current nanofabrication technologies and can be epitaxially grown on silicon [17], c-sapphire, and magnesium oxide [16]. Hence, in our simulations, we have considered the resonator to be made of titanium nitride (TiN) and the substrate to be MgO. We have performed simulations with two well-known magnetic PMA materials Bi-substituted iron garnet (BIG) [18] and Gallodinium Iron Cobalt (GdFeCo) [19]. Whereas we recognize that the magnetic properties of current BIG materials need to be improved to meet the thermal stability requirements for PMA nanomagnets, our goal in this paper in using BIG is to emphasize the potential improvements that can be obtained by employing dielectric nanomagnets.

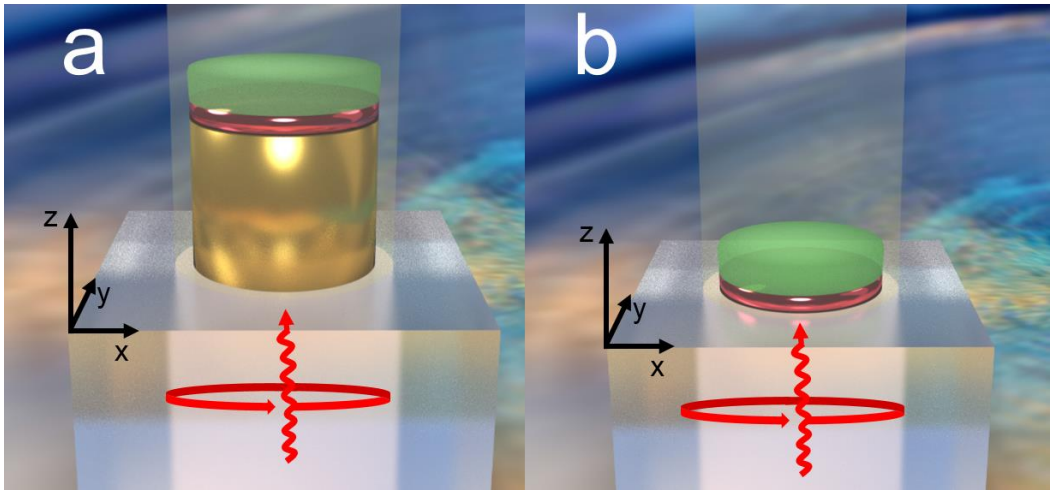

Fig. 1: (a) Schematic of proposed design. The substrate used is MgO. Each nanodisk consists of a plasmonic antenna (yellow), with a thin magnetic layer (purple) with a capping layer (green) on top. (b) Schematic of only a nanomagnet with the capping layer. In both figures, the red circular arrow at the bottom indicates that the illumination is with circularly polarized light and the curly red arrow indicates the direction of incidence.

As capping layer we selected $Si_3N_4$. The stack diameter is 50nm with the thickness of the TiN and the $Si_3N_4$ being 30nm and 20nm respectively. The thickness of the magnetic layer is kept constant at 10nm for both types of magnetic materials, BIG and GdFeCo. The choice of the stack diameter is selected to illustrate the key findings from our work. Later in the paper, we study the effect of increasing/decreasing the nanodisk diameter on the magnitude of the opto-magnetic field $H_{OM}$. The optical properties of TiN depend on the material thickness and a 30nm nanodisk exhibits good metallic properties in the visible and NIR [20,21]. The choice of $Si_3N_4$ as the capping layer material and its thickness (20nm) allows plasmonic excitations in the visible/NIR wavelength region of interest. In all simulations, we consider normal incidence illumination by circularly polarized light through the MgO substrate, as shown in Fig. 1a. The excitation wavelength is chosen to match the TiN nanodisk plasmonic resonance. This is determined to be 710nm for the material dimensions and optical constants employed. The illumination intensity is chosen as 1mJ/cm² in order to compare our opto-magnetic field magnitude estimates with those of ref. [5]. In Fig. 1b, we show a schematic of the same structure as Fig. 1a but without the plasmonic resonator. We refer to the stack in Fig. 1a containing the plasmonic resonator as magneto-plasmonic stack (MPS), whereas that without the resonator is identified as non-plasmonic stack (NPS). In the following sections we report almost an order of magnitude increment in the magnitude of the opto-magnetic field when the interaction with the nanomagnet is via the plasmonic resonator in the MPS sample. This significant increment is attributed to the near-field enhancement induced by the plasmonic resonator exhibiting Surface Plasmon Resonances (SPR). This enhancement indicates efficient coupling of electromagnetic radiation with an MPS structure as compared to direct illumination of nanomagnets as it is done in experiments reported thus far on all-optical-magnetization switching.

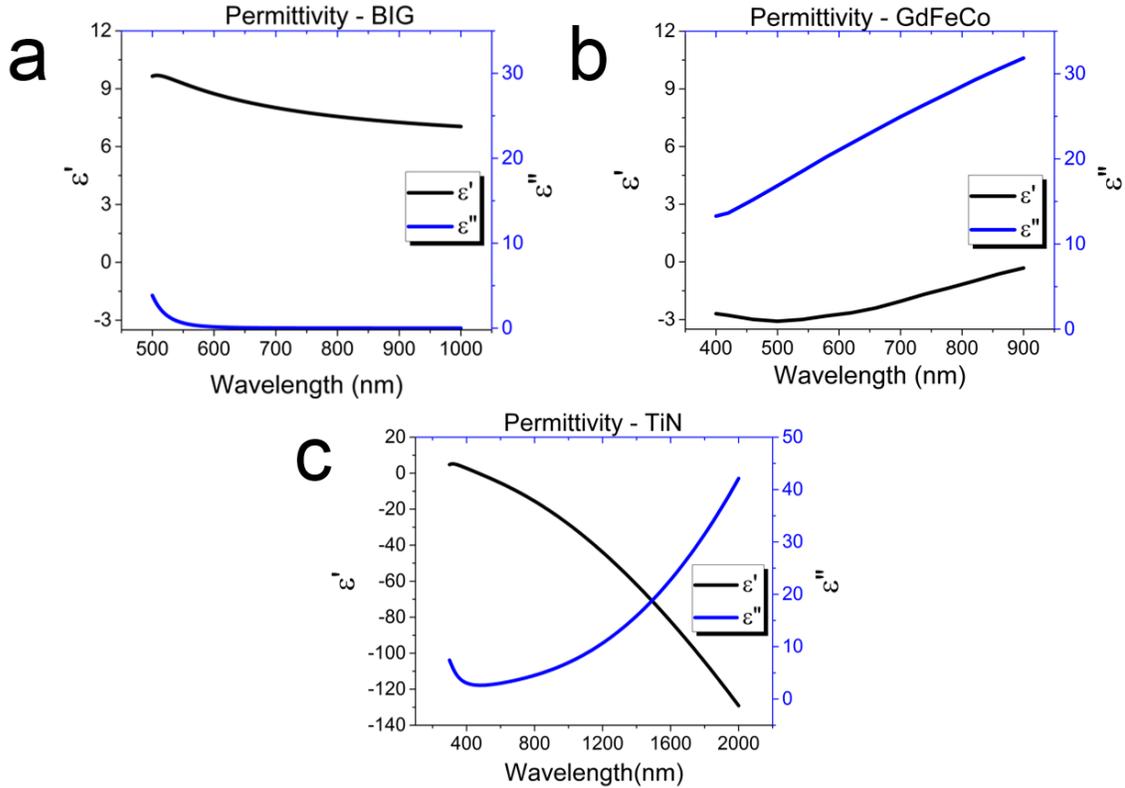

Figure 2: (a) Permittivity of Bismuth Iron Garnet (BIG) obtained from ref. [22]. (b) Permittivity of GdFeCo obtained from ref. [23]. (c) Permittivity of plasmonic TiN on MgO experimentally measured in our laboratory from spectroscopic ellipsometry measurements.

Figures 2a and 2b show the optical properties of BIG and GdFeCo in the visible and near-infrared regions. Figure 2c presents the optical properties of plasmonic TiN derived from spectroscopic ellipsometry measurements of sputter-deposited 30nm thick TiN thin films on MgO. Comparison of Fig. 2a and 2b indicate that at 710nm, BIG behaves as a lossless dielectric ($\varepsilon''$ is zero) whereas GdFeCo is a lossy metal. This results in plasmonic resonances in the BIG-MPS sample at the TiN-BIG interface, in contrast to the GdFeCo-MPS. There both GdFeCo and TiN are metallic in nature, the former being a relatively weaker metal (lower value of negative $\varepsilon'$). In this case we observe weak plasmonic resonances in the GdFeCo MPS sample at the GdFeCo-$Si_3N_4$ interface. More details are discussed in the following section.

## 3. RESULTS AND DISCUSSIONS

In our simulations, we used a finite-element frequency domain solver using COMSOL software [24]. For all simulations the illumination was with circularly polarized light at normal incidence. First we considered illumination of individual BIG-TiN MPS and NPS nanodisks separately and studied the electric field distribution characteristic of the plasmon resonance at 710nm wavelength. Utilizing Eq. (1), we derived the opto-magnetic field $H_{OM}$. The magnetic susceptibility $\beta$ is calculated using Eq. (2) with experimentally reported values of $\theta_F$ [25] and $M_0$ [26] for BIG. In Fig. 3a, we compare the magnitude of the z component of the opto-magnetic field component along the x-axis in BIG-TiN and BIG-MgO interface for MPS and NPS configurations respectively. One can clearly see a 10-fold enhancement in the z-component of the opto-magnetic field in the MPS sample compared to the NPS sample. In Fig. 3b, we show the variation of the opto-magnetic field in the BIG MPS sample (50nm diameter, 10 nm thick) as a function of wavelength. The data is taken at the center of the MPS stack at the BIG-TiN interface. The enhancement in $H_{OM,z}$ due to the plasmon resonance at 710nm wavelength is apparent. For additional details, the reader is referred to Fig. S1 in the Supplementary where we show the electric field amplitude color plots for light as a function of wavelength.

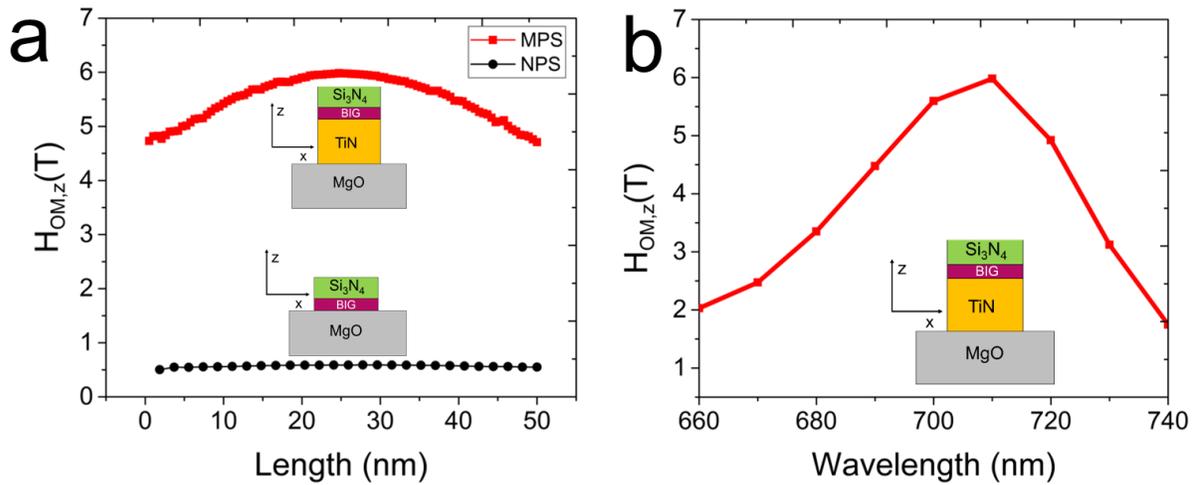

Figure 3: a. Comparison of the z-component of the opto-magnetic field intensity along the x-axis of BIG-TiN interface for a 10nm thick BIG layer in the MPS (nanomagnet with TiN resonator) and NPS (only nanomagnet) sample. Illumination is with circularly polarized light of intensity 1mJ/cm$^2$ at 710nm wavelength under normal incidence. b. Wavelength dependence of the z-component of the opto-magnetic field for the MPS sample (50 nm diameter) at the stack center at the TiN-BIG interface. Inset: Schematic of structure.

Next we present detailed electromagnetic analysis of the BIG MPS and NPS stacks when illuminated with circularly polarized light at 710nm. It is well known that when a plasmonic nanodisk is illuminated with circularly polarized light it generates a rotating dipole moment that is associated with the plasmonic resonance [27]. Figures.4a and 4c show the electric field components along the x-axis of the BIG-TiN and BIG-MgO interface for the MPS and NPS stack respectively. Comparing Fig. 4b and 4d we can clearly see the near-field enhancement for the MPS sample as compared to the NPS sample due to the plasmon resonance. Such a field enhancement is responsible for the increase in the opto-magnetic field of the MPS sample as noted earlier. A noticeable feature of the plasmon resonance is that it introduces an out-of-plane component of the electric field $E_z$ whereas the incident light has no such z-component. This is a well-known feature of the plasmon resonance [8]. For calculating the opto-magnetic field we have used only the out-of-plane component of the magnetic susceptibility $\beta$ as reported in literature. Thus, the opto-magnetic field reported in the figures, corresponds only to the z-component of $H_{OM}$. We note that due to the presence of additional components of the electric field, plasmon resonances can in principle generate in-plane components of the opto-magnetic field $H_{OM,x/y}$. The amplitudes of said components depend on the magnitude of $\beta$ in the x,y plane. Such an in-plane component of $H_{OM}$ can lead to more efficient switching of magnets with canted magnetization.

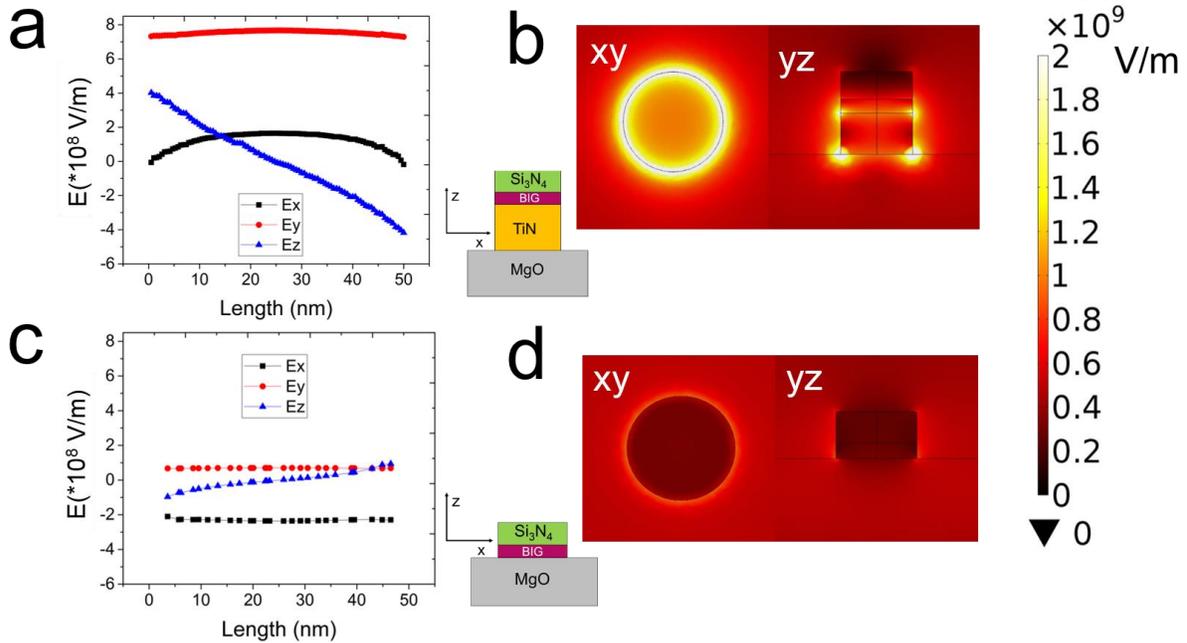

Fig. 4: (a) Electric field components along the x-axis of BIG-TiN interface for a 10nm BIG layer in the MPS sample. (b) Electric field intensity plot along the xy and yz plane for the 10nm BIG-TiN MPS structure under illumination with circularly polarized light of intensity 1mJ/cm$^2$ at 710nm wavelength. Top inset: Schematic of the vertical cross-section of BIG-TiN MPS sample (c) Electric field components along the x-axis of BIG-MgO interface for a 10nm BIG layer in the NPS sample. (d) Electric field intensity plot along the xy and yz plane for the 10nm BIG NPS structure under illumination with circularly polarized light of intensity 1mJ/cm$^2$ at 710nm wavelength. Bottom inset: Schematic of vertical cross-section of BIG-MgO NPS structure.

In Fig. 5a, we present the results on the opto-magnetic field $H_{OM}$ obtained for the MPS and NPS samples under illumination of circularly polarized light at 710nm wavelength when replacing the BIG with a 10nm thick GdFeCo nanomagnet. The magnetic susceptibility $\beta$ is obtained from [5] as $1.7*10^{-6}$ m/A. As noted earlier in Fig. 2b, GdFeCo is metallic in the visible/near-infrared region of the electromagnetic spectrum. Therefore in this case no plasmonic field enhancement at the GdFeCo-TiN interface is observed as both layers are metallic. In contrast, a plasmonic near-field enhancement at the GdFeCo-Si$_3$N$_4$ interface is observed in the MPS sample, although said enhancement is weaker. This is clearly shown in Fig. 5b and 5c where we plot the electric field intensity for the GdFeCo MPS and NPS sample respectively. The reason for the weaker field enhancement is because compared to TiN at 710nm wavelength, GdFeCo is a more lossy (higher $\varepsilon''$) and weaker (lower negative $\varepsilon'$) metal. The plasmonic resonance is mostly dominated by GdFeCo, with the TiN contribution to the plasmon resonance being significantly reduced. Hence the MPS structure with GdFeCo does not exhibit large opto-magnetic field enhancement compared to the NPS sample. We note an opto-magnetic field enhancement of only 3.4 in the MPS sample compared to the NPS sample which is less than what we observed for BIG due to aforementioned reasons. Although GdFeCo exhibits PMA for thicknesses > 10nm, the use of thicker layers would result in reduction of the contributions of the TiN plasmonic resonator on the opto-magnetic field enhancement.

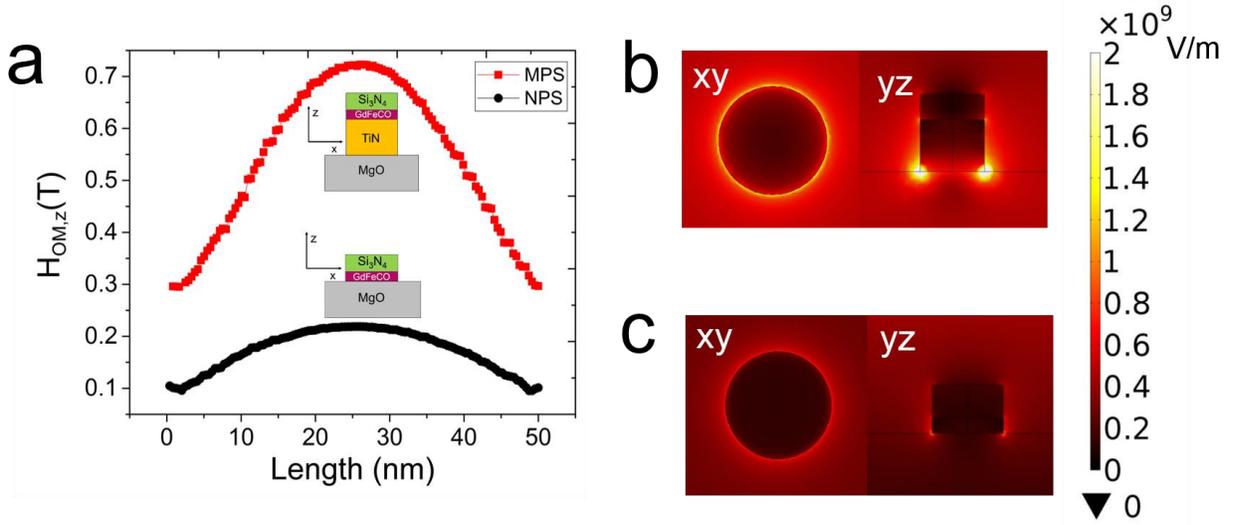

Fig. 5: a. Comparison of the z-component of the opto-magnetic field intensity along the x-axis of GdFeCo-$Si_3N_4$ interface for a 10nm GdFeCo layer MPS (nanomagnet with TiN resonator) and NPS (only nanomagnet) sample. b. Electric field intensity plots along the xy(GdFeCo-$Si_3N_4$) and yz plane for the MPS(nanomagnet with TiN resonator) sample. c. Electric field intensity plots along the xy(GdFeCo-$Si_3N_4$) and yz plane for the NPS(only nanomagnet) sample. Illumination is with circularly polarized light of intensity 1mJ/$cm^2$ at 710nm wavelength under normal incidence.

We next consider the dependence of the opto-magnetic field enhancement on the MPS diameter size. It is well-known that plasmonic resonances in metal nanodisks strongly depend on their size. A change in size can alter the strength of the plasmon resonance. Figure 6 presents the simulation results for an MPS sample with 10nm BIG. In Fig. 6a, we plot the opto-magnetic field z-component along the x-axis in the x-y plane of the BIG-TiN interface for a 20nm diameter stack. A clear enhancement in $H_{OM,z}$ of ~3x is observed compared to free space excitation (Fig. 5b). However, a comparison of Fig. 3 and Fig. 6a indicate that a reduction of the stack diameter of the MPS sample reduces the plasmonic enhancement in $H_{OM}$ from 10 to 3. Also the nature of the opto-magnetic field along the diameter is different. We believe that this is due to the weaker nature of the plasmon resonance for the sample with reduced diameter. Comparing Fig. 4a and Fig 6c, one can clearly see that the y-component of the electric field is reduced upon reducing the diameter. This leads to a reduction in $H_{OM,z}$. Also the variation of the y-component of the electric field of light along the stack diameter is slightly different upon reducing the diameter of the stack. Consequently there is a change in the way $H_{OM,z}$ varies along the diameter when the stack diameter is reduced. Nevertheless, the plasmonic enhancement in $H_{OM,z}$ still persists. This illustrates that plasmon-mediated magnetic switching is compatible with dimensionality reduction of nanomagnets for ultra-high density storage and it points to opportunities to increase the effect by matching the excitation light source to the plasmon resonance.

Table 1 presents a summary of the obtained simulation results. The Figure-of-Merit *FOM* is calculated using the following formula:

$$\text{FOM} = \frac{H_{OM,z(MPS)}}{H_{OM,z(NPS)}} \qquad (3)$$

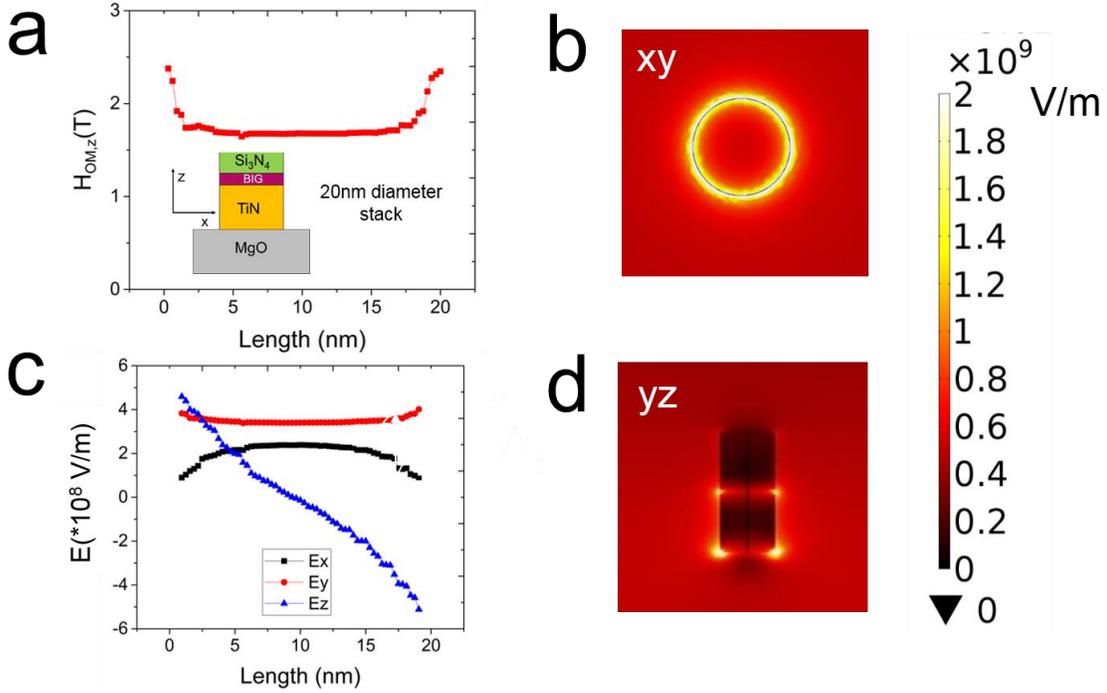

Fig. 6: a. Opto-magnetic field intensity along the x-axis of BIG-TiN interface for MPS structure with 20nm diameter. Inset: Schematic of structure. b. Electric field intensity plot at TiN–BIG interface. c. Electric field components along the x-axis of the TiN-BIG interface. d. Electric field intensity plot along the yz plane of the MPS stack. Illumination is with circularly polarized light of intensity 1mJ/cm$^2$ at 710nm wavelength.

We note that that in our simulations we neglect the optical anisotropy of both GdFeCo and BIG magnetic layers. This is because the off-diagonal terms of the permittivity tensor for both materials, obtained from the product of $\beta$ and $M_0$ are at least an order of magnitude lower than the diagonal terms as shown in Fig. 2a and 2b.

Table 1: Summary of Results

| Magnetic Material | n | $\theta_F$ (deg.) | $M_0$ (A/m) | $\beta$ (m/A) | FOM |
|---|---|---|---|---|---|
| BIG | 2.82 | 0.027[25] | 5.5*10$^4$[26] | 5.5*10$^{-7}$ | 10 |
| GdFeCo | 3.42 | 0.75[5] | 1*10$^5$[5] | 1.7*10$^{-6}$[5] | 3.4 |

## 4. CONCLUSION

We report on a numerical study of surface plasmon-induced opto-magnetic fields for manipulation of the magnetic orientation of nanomagnets in on-chip, CMOS-compatible structures. We have used plasmonic titanium nitride and two commonly used PMA materials BIG and GdFeCo as nanomagnets to provide a fully CMOS-compatible material platform for the realization of future on-chip hybrid nanoplasmonic-magnetic devices. Our results show that surface plasmon resonances can lead to significant near-field enhancement and thereby much higher opto-magnetic field $H_{OM}$ compared to free space excitation. We further show that the optical properties of the magnetic material play a crucial role in determining the strength of the plasmon resonance and hence an optimal choice of material is necessary in order to achieve the highest possible $H_{OM}$. We also note the possibility of having in-plane components of the opto-magnetic field $H_{OM}$ with plasmon resonance. Thus, the proposed approach could potentially enable future on-chip plasmonic-magnetic devices for spintronics and memory.

## 5. ACKNOWLEDGEMENT

The authors would like to acknowledge Dr. Kuntal Roy for useful suggestions and discussions. The authors would also like to acknowledge support from the Office of Naval Research Grant No. N00014-16-1-3003.

# Surface-plasmon opto-magnetic field enhancement for all-optical magnetization switching

A. Dutta[a,b*], A. V. Kildishev[a,b], V. M. Shalaev[a,b], A. Boltasseva[a,b], E. E. Marinero[a,b,c]

[a]School of Electrical and Computer Engineering, [b] Birck Nanotechnology Center, [c]School of Materials Engineering, Purdue University, West Lafayette IN, USA 47907

*Corresponding authors: dutta6@purdue.edu, eemarinero@purdue.edu

## SUPPLEMENTARY

In Fig. S1 we present some further results on the 50nm diameter BIG-TiN MPS sample simulation. Fig. S1a shows the opto-magnetic field $H_{OM,Z}$ at the center of the stack at the BIG-TiN interface as a function of illumination wavelength. It is the same as Fig. 3b in the main paper and is repeated here for the sake of convenience. In Fig. S1b, we look at the electric field amplitude color maps as a function of the incident circularly-polarized wavelength. The intensity of the incident field is 1mJ/cm$^2$ for all the wavelengths. It is clear from the color maps in Fig. S1b, that as we approach 710nm wavelength, the near field intensity reaches a maxima. This in turn leads to the highest opto-magnetic field $H_{OM,Z}$ at 710nm wavelength as shown in Fig. S1a and Fig. 3b. As we cross 710nm towards higher wavelengths, then the near field intensity again starts to fall. This allows us to conclude the presence of a surface plasmon resonance for this structure at 710nm wavelength under illumination by circularly-polarized light.

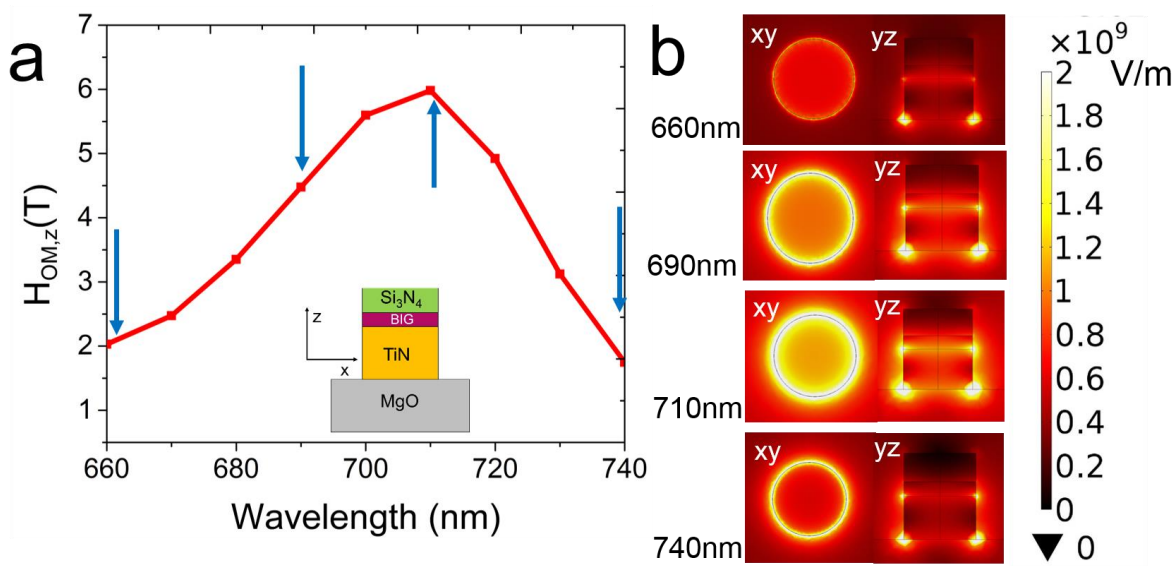

Fig. S1: a. Wavelength dependence of the z-component of the opto-magnetic field for the 50nm diameter MPS sample at the stack center at the TiN-BIG interface. Inset: Schematic of the structure. Blue arrows correspond to the wavelengths for which the electric field amplitude of light is plotted in b. b. Electric field intensity plot along the xy and yz plane for the 10nm BIG-TiN MPS structure under illumination with circularly polarized light of intensity 1mJ/cm$^2$. The wavelength of excitation is shown at the bottom left corner (xy plane refers to the BIG-TiN interface).